\documentclass[prl, reprint, superscriptaddress, showpacs, bibliography]{revtex4-1}
\usepackage[dvips]{graphicx}
\usepackage{color}
\usepackage{amsmath}
\usepackage{amssymb}
\usepackage[breaklinks]{hyperref}
\usepackage[all]{hypcap}

\begin{document}

\title{Nuclear Magnetic Resonance Reveals Disordered Level-Crossing Physics in the Bose-Glass Regime of the Br-doped Ni(Cl$_{1-x}$Br$_x$)$_2$-4SC(NH$_2$)$_2$ Compound\\ at a High Magnetic Field}

\author{Anna Orlova}
\affiliation{Laboratoire National des Champs Magn\'etiques Intenses, LNCMI-CNRS (UPR3228), \\ EMFL, UGA, UPS, and INSA, Bo\^{i}te Postale 166, 38042, Grenoble Cedex 9, France}

\author{R\'{e}mi Blinder}
\altaffiliation[Present address: ]{CEA, INAC, SPRAM (UMR5819 CEA/CNRS/UJF), 17 rue des Martyrs, 38054 Grenoble cedex 9, France}
\affiliation{Laboratoire National des Champs Magn\'etiques Intenses, LNCMI-CNRS (UPR3228), \\ EMFL, UGA, UPS, and INSA, Bo\^{i}te Postale 166, 38042, Grenoble Cedex 9, France}

\author{Edwin Kermarrec}
\altaffiliation[Present address: ]{Laboratoire de Physique des Solides, Universit\'{e} Paris-Sud, UMR CNRS 8502, 91405 Orsay, France}
\affiliation{Laboratoire National des Champs Magn\'etiques Intenses, LNCMI-CNRS (UPR3228), \\ EMFL, UGA, UPS, and INSA, Bo\^{i}te Postale 166, 38042, Grenoble Cedex 9, France}

\author{Maxime Dupont}
\affiliation{Laboratoire de Physique Th\'{e}orique, IRSAMC, Universit\'{e} de Toulouse, CNRS, 31062 Toulouse, France}

\author{Nicolas Laflorencie}
\affiliation{Laboratoire de Physique Th\'{e}orique, IRSAMC, Universit\'{e} de Toulouse, CNRS, 31062 Toulouse, France}

\author{Sylvain Capponi}
\affiliation{Laboratoire de Physique Th\'{e}orique, IRSAMC, Universit\'{e} de Toulouse, CNRS, 31062 Toulouse, France}

\author{Hadrien Mayaffre}
\affiliation{Laboratoire National des Champs Magn\'etiques Intenses, LNCMI-CNRS (UPR3228), \\ EMFL, UGA, UPS, and INSA, Bo\^{i}te Postale 166, 38042, Grenoble Cedex 9, France}

\author{Claude Berthier}
\affiliation{Laboratoire National des Champs Magn\'etiques Intenses, LNCMI-CNRS (UPR3228), \\ EMFL, UGA, UPS, and INSA, Bo\^{i}te Postale 166, 38042, Grenoble Cedex 9, France}

\author{Armando Paduan-Filho}
\affiliation{Instituto de F\'isica, Universidade de S{\~a}o Paulo, 05315-970 S{\~a}o Paulo, Brazil}

\author{Mladen Horvati\'c}
\email{mladen.horvatic@lncmi.cnrs.fr}
\affiliation{Laboratoire National des Champs Magn\'etiques Intenses, LNCMI-CNRS (UPR3228), \\ EMFL, UGA, UPS, and INSA, Bo\^{i}te Postale 166, 38042, Grenoble Cedex 9, France}

\begin{abstract}
By measuring the nuclear magnetic resonance (NMR) $T_1^{-1}$ relaxation rate in the Br (bond) doped DTN compound, Ni(Cl$_{1-x}$Br$_x$)$_2$-4SC(NH$_2$)$_2$ (DTNX), we show that the low-energy spin dynamics of its high magnetic field ``Bose-glass'' regime is dominated by a strong peak of spin fluctuations found at the nearly doping-independent position $H^* \cong 13.6$~T. From its temperature and field dependence we conclude that this corresponds to a level crossing of the energy levels related to the doping-induced impurity states. Observation of the local NMR signal from the spin adjacent to the doped Br allowed us to fully characterize this impurity state. We have thus quantified a microscopic theoretical model that paves the way to better understanding of the Bose-glass physics in DTNX, as revealed in the related theoretical study [M. Dupont, S. Capponi, and N. Laflorencie, \href{https://doi.org/10.1103/PhysRevLett.118.067204}{Phys. Rev. Lett. \textbf{118}, 067204 (2017)}, \href{https://arxiv.org/abs/1610.05136}{arXiv:1610.05136}].\end{abstract}

\date{\today}




\doi{10.1103/PhysRevLett.118.067203}

\maketitle

The NiCl$_2$-4SC(NH$_2$)$_2$ (DTN) compound \cite{Paduan81}, consisting of weakly coupled chains of $S = 1$ (Ni-ion) spins with an easy-plane single-ion anisotropy ($D$), is one of the most studied quantum spin materials \cite{Zapf14}. Between the two critical magnetic fields $H_{\textrm{c1}}$ and $H_{\textrm{c2}}$, it presents a magnetic-field-induced low-temperature ($T$) 3D-ordered phase, described as a Bose-Einstein condensate (BEC) \cite{Giamarchi99,Nikuni00,Giamarchi08,Zapf14}. DTN is particularly convenient for studying this phase; for a magnetic field ($H$) applied along the chain $c$ axis, its (body-centered) tetragonal symmetry \cite{Figgis86} ensures the required axial symmetry of the spin Hamiltonian with respect to $H$. The values of its exchange couplings and $D$ ($J_c/k_B = 2.2$~K, $J_{a,b}/k_B = 0.18$~K, $D/k_B = 8.9$~K) \cite{Zvyagin07, Tsyrulin13} make the BEC phase easily accessible, with $H_{\textrm{c2}} = 12.32$~T \cite{Paduan04,Zapf06,Mukhopadhyay12} and the phase transition temperature $T_\textrm{c}$ below $T_\textrm{cmax} = 1.2$~K. The system can be reasonably considered as quasi-one-dimensional (1D), with $J_{a,b}/J_c = 0.08$.

Br-doped DTN, Ni(Cl$_{1-x}$Br$_x$)$_2$-4SC(NH$_2$)$_2$ (DTNX), allows studying the effect of a bond disorder, which may lead to the appearance of a localized Bose-glass (BG) phases adjacent to the (now inhomogeneous) BEC phase \cite{Zheludev13}, as suggested from the thermodynamic measurements \cite{Yu12}. The BG state, first discussed for quantum wires \cite{Giamarchi88} and superfluid $^4$He absorbed in porous media \cite{Fisher89, Reppy92}, remains elusive, with only a few experimental examples \cite{Fallani07,Malpuech07,Roux08,Sacepe11}, particularly rare for condensed-matter systems in the thermodynamic limit \cite{Hong10}, such as DTNX \cite{Yu12}.

We present here the first microscopic information on the high-field ($H > H_{\textrm{c2}}$) disordered state in DTNX, where the low-energy spin fluctuations (dynamics) are measured by $^1$H and $^{14}$N nuclear spin-lattice relaxation rate ($T_1^{-1}$), while the NMR spectra revealed the local spin polarization \cite{Horvatic02}. As compared to pure DTN, the main feature of spin dynamics in DTNX is a peak of $T_1^{-1}$ appearing at $H^* \cong 13.6$~T independently of the doping level. This is attributed to the level crossing of single-particle states strongly localized at the doped-bond position, which is then somewhat distributed/disordered by the mutual interaction of these states. The disorder is seen by NMR as the inhomogeneous relaxation, relatively broad width of the $T_1^{-1}$ peak, and by the modification of the Arrhenius (gapped) $T_1^{-1}(T)$ dependence. From the NMR spectra we also determined the local spin-polarization value of the spin adjacent to the doping position, which allows us to determine the related local impurity exchange coupling ($J'$) and the impurity single-ion anisotropy ($D'$) \cite{Yu12, Dupont16}. We thereby fully characterize the impurity state created by doping, and find that it is strongly localized on impurity, having extremely short correlation lengths. Above $H_{\textrm{c2}}$, many-body physics is thus controlled by the effective pairwise interaction between such localized impurity states. This sheds a new light on the BG physics expected in this system above $H_{\textrm{c2}}$~\cite{Yu12}; while a BG regime implies the absence of any long-range order down to $T = 0$, in the related theoretical work \cite{Dupont16} it is shown that around $H^*$ the impurity states order at low $T$, thereby profoundly changing the phase diagram.

The present investigation of DTNX is based on previous NMR work on DTN samples \cite{Blinder15, Blinder16}, enabling precise knowledge of the hyperfine coupling tensors ($\textbf{A}$) that relate the spin polarization to the local magnetic field observed by the NMR frequency shift of $^1$H and $^{14}$N nuclei. Three different doping levels, $x = 4\%$, 9\%, and 13\%, of DTNX single crystals were studied in a $^4$He cryostat, with $H$~$\|$~$c$ within 1$^{\circ}$. The 4\% doped crystal was further studied at a lower temperature in a dilution refrigerator, with a fixed orientation where the $c$-axis tilt was 2.5$^{\circ}$.

\begin{figure}[t!]
\includegraphics[width=1.00\columnwidth,clip,angle=0]{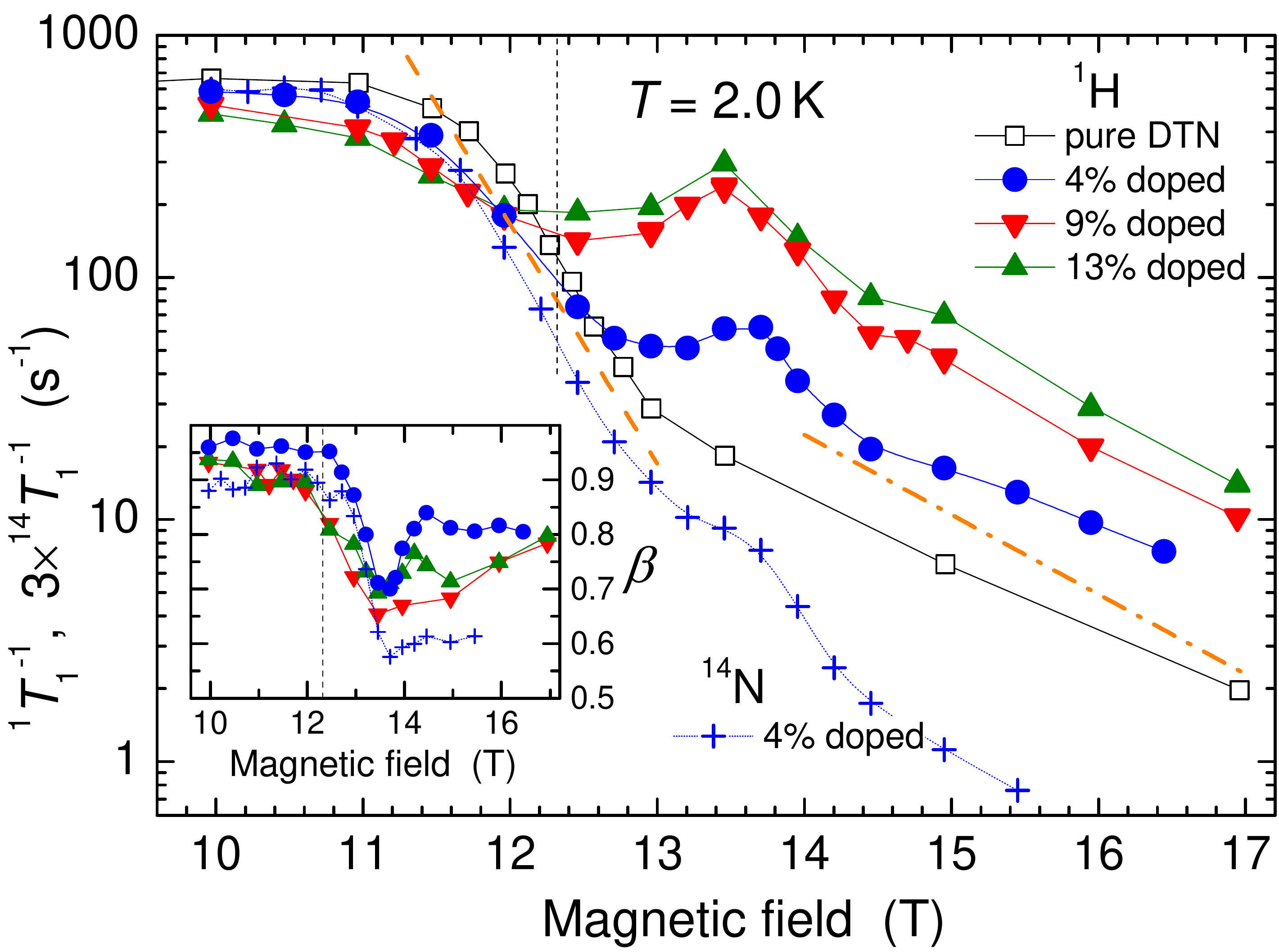}
\caption{Magnetic field dependence of $T_1^{-1}$ measured by $^{14}$N (crosses) and $^1$H (other symbols) NMR in DTN (from Ref.~\cite{Mukhopadhyay12}) and Br-doped DTNX at 2~K. Vertical dotted line denotes the $H_{\textrm{c2}} = 12.32$~T position in DTN. Orange dashed and dash-dotted lines show gapped behavior, $\textrm{e}^{-3\Delta/T}$ and $\textrm{e}^{-\Delta/T}$, respectively. Inset: exponents $\beta$ of the ``stretched exponentials'' used to fit the relaxation curves in DTNX.}\label{DTNXfig1}
\end{figure}

Figure~\ref{DTNXfig1} presents the doping dependence of $^1$H $T_1^{-1}$ taken at $T = 2.0$~K, which is well above $T_\textrm{cmax}$. The system is thus in the 1D Tomonaga-Luttinger liquid regime above the BEC phase \cite{Mukhopadhyay12}. Below $H_{\textrm{c2}}$ one thus observes high $T_1^{-1}$ values (spin fluctuations), which are nearly doping independent. Upon increasing field this is followed by the critical regime around $H_{\textrm{c2}}$, presenting for the pure DTN a strong decrease of spin fluctuations. This dependence is very close to what is predicted for the critical regime by the second order processes for the transverse spin fluctuations \cite{Orignac07},  $\langle$$S_+S_-\rangle \propto \textrm{e}^{-3\Delta/T}$, corresponding to triple of the gap value,  $\Delta = g_c\mu_B(H - H_{\textrm{c2}})/k_B$, that magnetic field opens for a spin-1 excitation, where the $g$-tensor value is $g_c = 2.26$ [7]. One tesla above $H_{\textrm{c2}}$ this strong decrease of $T_1^{-1}$ is slowed down as the longitudinal fluctuations become dominant; they are indeed expected to decrease much slower, directly reflecting a (single) gap opening,  $\langle$$S_zS_z\rangle \propto \textrm{e}^{-\Delta/T}$, which is close to what is observed in the $T_1^{-1}$ data at the high-field end for all doping values~\cite{gapped?}.

The most obvious effect of the doping is the appearance of a peak in the $^1$H $T_1^{-1}$ at 13.6~T, whose position is nearly independent of doping while its strength increases and then saturates above $x = 10\%$. This is accompanied by a strong spatial inhomogeneity of $T_1^{-1}$, observed as a stretched exponential relaxation of nuclear magnetization, $\propto \textrm{e}^{-(t/T_1)^{\beta}}$. The $\beta$ values reflect the spread of the distribution of local $T_1^{-1}$ values, decreasing from $\beta = 1$ for a homogeneous system, to reach a distribution width of one order of magnitude (on a logarithmic scale) already at $\beta = 0.74$ \cite{Johnston06, Mitrovic08, Blinder15}. Here the spatial inhomogeneity of local spin fluctuations develops in DTNX above $H_{\textrm{c2}}$, and is particularly strong at the $T_1^{-1}$ peak.

\begin{figure}[b!]
\includegraphics[width=1.00\columnwidth,clip,angle=0]{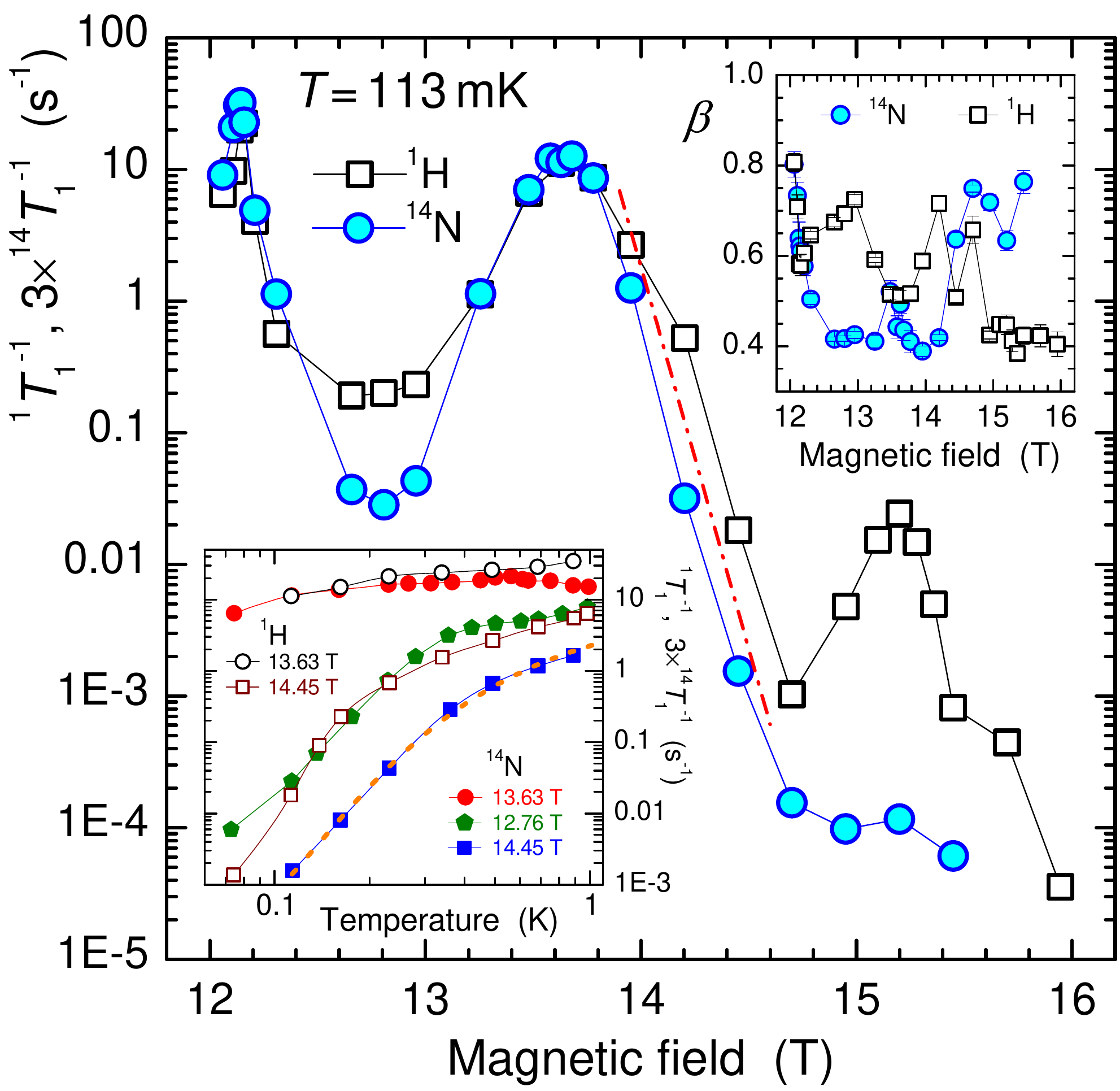}
\caption{Magnetic field dependence of $T_1^{-1}$ measured by $^1$H and $^{14}$N NMR in 4\% doped DTNX at 113~mK. Red dash-dotted line presents the $\textrm{e}^{-\Delta/T}$ gapped behavior. Upper inset: stretch exponents corresponding to the $T_1^{-1}$ data. Lower inset: temperature dependence of $T_1^{-1}$ at three characteristic field values. Orange dashed line is the ``distributed gap'' fit, see the text.}\label{DTNXfig2}
\end{figure}

\begin{figure}[t!]
\includegraphics[width=1.00\columnwidth,clip,angle=0]{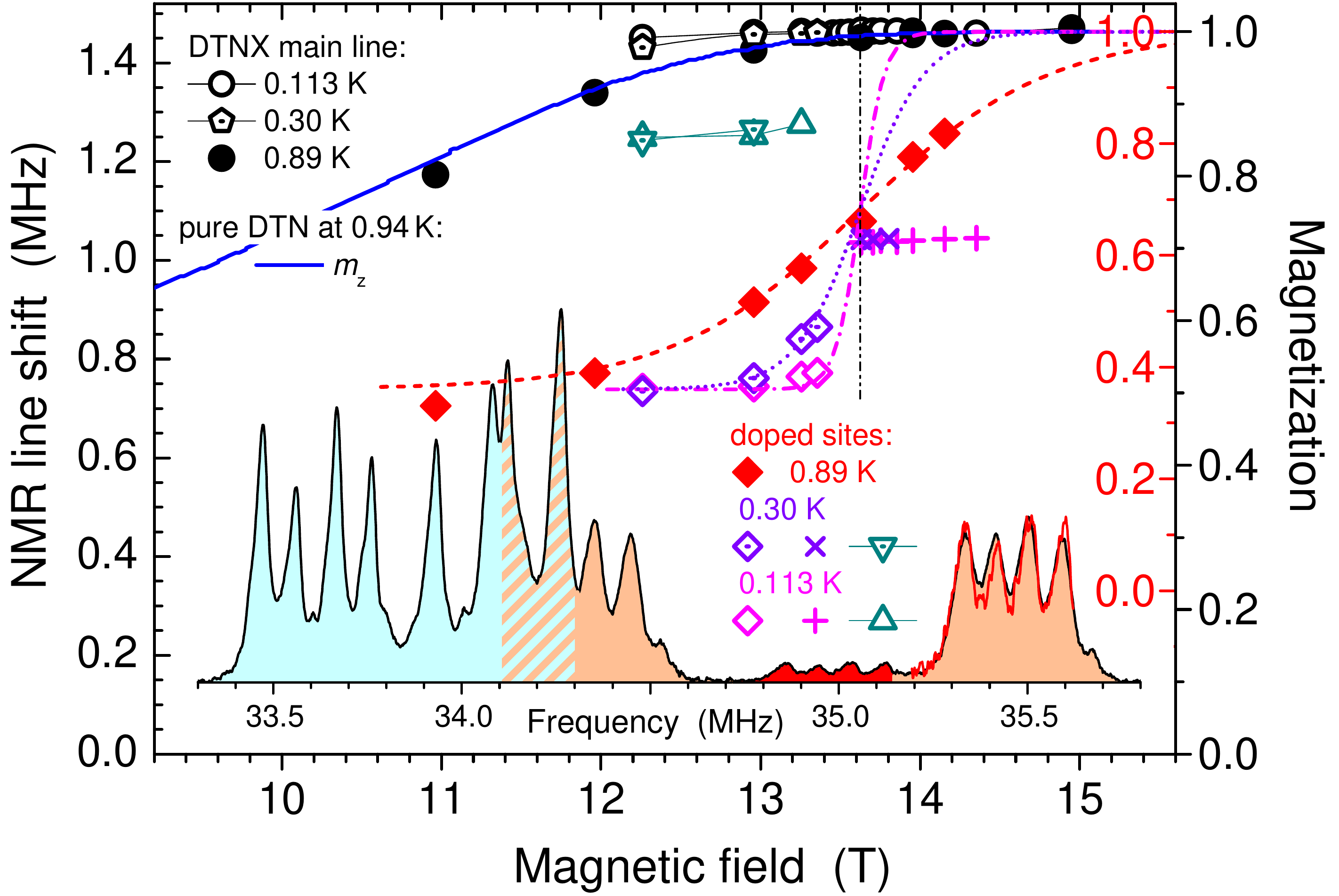}
\caption{$H$ dependence of the local magnetization (right scales) measured in 4\% DTNX by the NMR line position (left scale) of the high-frequency $^{14}$N regular line (circles and pentagons, outer right scale) and impurity line (other symbols, inner right scale for diamonds and crosses) at 0.89, 0.30 and 0.113~K. Blue solid line is the 0.94~K magnetization of pure DTN from Ref.~\cite{Paduan04}. Vertical dash-dot-dot line denotes the level-crossing value $H^*  =  13.63$~T. Other tanh-shape lines are the ``two-level fits'', see the text. Inset: $^{14}$N spectrum at 0.89~K and 10.96~T, where the colors (or gray hues) denote different contributions: light blue, brown and red, for the N(2), N(1) sites (including the overlap of the two sites), and the doped ``impurity'' N(1) site, respectively. By shifting and upscaling this latter contribution (red line), one can perfectly overlap the regular N(1) line shape.}\label{DTNXfig3}
\end{figure}

We remark that, in general, $T_1^{-1}$ is sensitive to both transverse and longitudinal spin fluctuations, through the diagonal and off-diagonal elements of its $\textbf{A}$ tensor, respectively. This is particularly valid for $^1$H $T_1^{-1}$ where all the elements of its $\textbf{A}$ tensor are of the same order of magnitude \cite{Blinder15, Blinder16}. In contrast to that, $T_1^{-1}$ measured on the high-frequency line of $^{14}$N (see inset to Fig.~\ref{DTNXfig3}), whose $\textbf{A}$ tensor is strongly dominated by its isotropic component, is strongly dominated by transverse spin fluctuations. The nature of spin fluctuations can thus be distinguished by comparing $T_1^{-1}$ measured on the two nuclei. For example, in Fig.~\ref{DTNXfig1} the $T_1^{-1}$ peak is barely visible in the $^{14}$N $T_1^{-1}$ data (measured only for the 4\% doped DTNX), meaning that at this temperature it dominantly reflects longitudinal spin fluctuations. Note that the $^{14}$N $T_1^{-1}$ values are multiplied by 3, in order to overlap the $^1$H and $^{14}$N data points below $H_{\textrm{c2}}$, where the transverse spin fluctuations are dominant \cite{Orignac07}.

Figure~\ref{DTNXfig2} presents the $T_1^{-1}$ data for the 4\% doped DTNX, both for $^{14}$N (closed symbols) and $^1$H (open symbols), taken at very low temperature, $T \cong  T_\textrm{cmax}/10$. There are two strong $T_1^{-1}$ peaks: a very sharp one at $H_{\textrm{c}}(113~\textrm{mK}) = 12.14$~T corresponds to critical fluctuations at the phase boundary of the ordered BEC phase, while the second one is at the same position as at a high temperature, $H^* = 13.63$~T. It has kept approximately the same width, which should thus be associated to an intrinsic disorder and not to a temperature effect. Since the transverse spin fluctuations are dominant at the BEC phase transition, the $^1$H and $^{14}$N data points overlap there. Unlike at high $T$, the same is now valid close to $H^*$, meaning that the nature of spin fluctuations has changed there. Outside of these two $T_1^{-1}$ peaks the $^1$H data points lay above the $^{14}$N data, reflecting the presence of longitudinal spin fluctuations. On the right-hand side of the $H^*$ peak, where the $T_1^{-1}(H)$ dependence is not affected by the critical regime related to $H_{\textrm{c2}}$, experimental points follow a gapped, $\propto \textrm{e}^{-\Delta/T}$, dependence~\cite{gapped?}. The inset to Fig.~\ref{DTNXfig2} shows that the $T_1^{-1}$ values taken at $H^*$ are nearly $T$ independent, while the $T_1^{-1}(T)$ dependence at 14.45~T is close to, but not exactly the same gapped behavior as observed in the $H$ dependence. Remembering that the relaxation is here spatially inhomogeneous, this can be easily modeled: for simplicity we assume that the corresponding distribution of the local gap values is Gaussian, centered at $\Delta_0$ and having the variance $\delta$. One can then analytically calculate the average gapped behavior, $\langle\textrm{e}^{-\Delta(r)/T}\rangle \propto \textrm{e}^{-\Delta_0/T}\textrm{e}^{\delta^2/(2T^2)}$. Fitting the 14.45~T $^{14}$N data to this dependence we estimate the distribution of $H^*$ values, $\delta = 0.2$~T, reflecting the disorder of the system. As $\delta$ is field independent, we also understand that the previously discussed $H$ dependence of $T_1^{-1}$, taken at fixed $T$, remains unmodified, simple gapped behavior.

We note that the peak in $T_1^{-1}$ is observed for all doping values at the same field value $H^*$ where the steplike increase is found in the magnetization measurements and associated to the number of dopants \cite{Yu12}. This magnetization step, together with the previously discussed (gapped) $H$ and $T$ dependences of $T_1^{-1}$ are then analogous to what is observed in the molecular antiferromagnetic rings, where a peak of $T_1^{-1}$ appears at each crossing of molecular levels \cite{Julien99, Micotti05}. In DTNX the peak should then be associated to crossing of levels of the impurity states localized on Br dopants. As the distance, and thus the mutual interaction between these states are (randomly) distributed, we speak of disordered level-crossing physics.

\begin{figure}[t!]
\includegraphics[width=1.00\columnwidth,clip,angle=0]{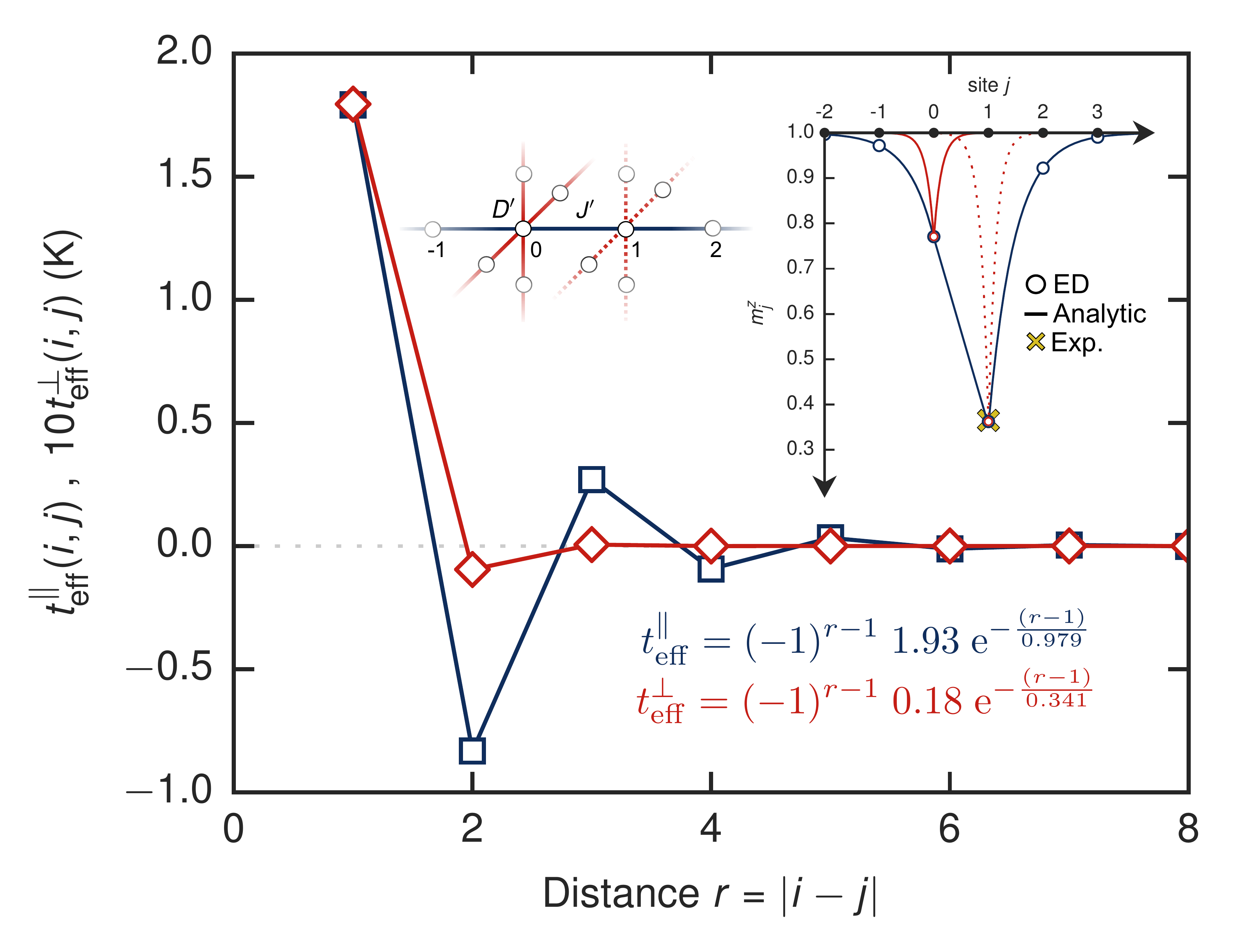}
\caption{Theoretical (ED) results for the effective pairwise coupling between impurity states as a function of their relative separation $r$. Both longitudinal $t_\mathrm{eff}^{\parallel}$ (squares) and transverse $t_\mathrm{eff}^{\perp}$ (diamonds) couplings display a staggered exponential decay. Left inset: schematic picture of a doped bond. Right inset: ED (circles), analytical (lines), and experimental (NMR, cross) values of local spin (de)polarizations, close to a doped bond, along the chain (blue) and perpendicular to it (red), for a depolarized single impurity ($H_{\textrm{c2}} < H < H^*$).}\label{DTNXfig4}
\end{figure}

The final confirmation of this scenario comes from the observation of the local spin value of one of the two impurity sites linked by the Br-doped bond. Indeed, in the $^{14}$N NMR spectrum shown in Fig.~\ref{DTNXfig3}, inside the gap produced by the main quadrupolar splitting of the high-frequency line ($^{14}$N being a spin-1 nucleus), one can clearly observe a small signal whose shape closely mimics the 4 peaks of the main line (where this fine-split quartet is due to the 2.5$^{\circ}$ tilt of the sample). This means that we observe the impurity spin site at position ``1'' in Fig.~\ref{DTNXfig4}, having unperturbed $D$ value. The intensity ratio of this impurity signal and the main line is indeed close to the expected value, $2x/(1 - 2x) = 0.09$ \cite{note}. We can thus separately measure the local polarizations of the impurity site and of the main regular sites away from the doped bonds. These latter are practically unaffected by doping and closely follow the magnetization of the pure DTN (Fig.~\ref{DTNXfig3}), defining thus the corresponding (global) hyperfine coupling $\textrm{A}_{cc}(q = 0)$. For a simple two-level crossing, the $H$ dependence of the impurity spin polarization has a Brillouin-like behavior, starting at $m_{\textrm{low}}$ and saturating at 1: $m_1(H,T) = \scriptsize{\frac{1-m_{\textrm{low}}}{2}} \tanh \scriptsize{\frac{g_c\mu_B(H-H^*)}{2k_BT}} + \scriptsize{\frac{1+m_{\textrm{low}}}{2}}$. As much as the NMR line position of the weak impurity signal could be resolved from the intense regular NMR lines, its position (diamond symbols in Fig.~\ref{DTNXfig3}) is indeed following this prediction. However, we can expect that the simple model of independent impurities will be affected by disorder close to $H^*$. Indeed, at low $T$ and close to $H^*$ we observed complex bicomponent spectra, and above $H^*$ the impurity signal was screened by a weak $H$-independent signal, denoted by crosses in Fig.~\ref{DTNXfig3}.

Extracting the value of the local spin  polarization from the frequency of the impurity line is nontrivial. The global $\textrm{A}_{cc}(q = 0)$ value, determined from saturated magnetization of the regular NMR line, consists of two unknown local couplings of $^{14}$N nucleus to the two closest (Ni) spins, and of known dipolar coupling to spins that are further away: $\textrm{A}_{cc}(q = 0) = \textrm{A}_1 + \textrm{A}_2 + \textrm{A}_\textrm{dip}$. Precisely these unknown local couplings are active in measurements of individual, local spin values. If $\textrm{A}_1$ and $\textrm{A}_2$ are comparable but unequal, a completely localized spin depolarization, ($1 - m_{\textrm{low}}$), will be seen as \textit{two} distinct NMR lines, corresponding to the two different neighboring $^{14}$N nuclei. We could identify the second NMR signal of the impurity site, shown by triangles in Fig.~\ref{DTNXfig3}, and thus completely determine all the couplings: $\textrm{A}_1 = 77.4\%$, $\textrm{A}_2 = 22.6\%$, $\textrm{A}_\textrm{dip} = 0.03\%$, all relative to $\textrm{A}_{cc}(q = 0) = 1.463$~MHz/spin-1. This defines the local polarization scale (right inner scale in Fig.~\ref{DTNXfig3}) and the local low-$T$ (de)polarization value $m_{\textrm{low}} = 0.365(5)$.

This $m_{\textrm{low}}$ value, together with $H^*$ determined from $^{14}$N $T_1^{-1}$ of Fig.~\ref{DTNXfig2}, enable us to directly determine the doping-induced, locally modified parameters $J'$ and $D'$. We further observe that $H^*$ is doping independent up to a very high percentage of doped bonds (2$x = 26\%$) \cite{note}, suggesting that these impurity states are strongly localized, and that a single-impurity model may capture most of the physics. Indeed, a level crossing naturally arises from a simple model describing a single doped (and thus more strongly coupled) $S = 1$ dimer embedded in its (mean-field, less coupled) clean environment, leading to qualitatively reasonable results. Furthermore, from this model we learn that one of these impurity states is precisely the localized state observed by neutron diffraction at $H = 0$ above the magnon band \cite{Yu15}, and that further localized states that can be seen in Fig.~7 of Ref.~\cite{Yu15}, slightly below the top and below the center of the band, belong to the same multiplet of impurity states.

Including quantum dynamics provides crucial information on the (de)localization of the spin around the impurities \cite{Dupont16}. For a single impurity, this can be studied both analytically or numerically by exact diagonalization (ED). Using $H^*$ and $m_{\textrm{low}}$ values we precisely determine the local parameters of a doped dimer, yielding $J' = 2.42J_c$ and $D' = 0.36D$~\cite{noteyu}. The magnetization profile of the $S^z_{\textrm{tot}} = 1$ impurity state displays an exponential decay, $1 - m_j^z \propto \textrm{e}^{-|j|/\xi}$, with a very short localization length $\xi_c \simeq 0.48$ (in lattice units) along the chains and $\xi_{a,b} \simeq 0.17$ in the transverse directions (inset of Fig.~\ref{DTNXfig4}), clearly demonstrating strong localization of the spin density on the doped dimer.

Importantly, such localized impurity states do not always behave as decoupled free spins. They might get correlated at low enough temperature as many-body effects are expected to arise from a distance ($r$) dependent pairwise coupling between randomly located impurities, although it should rapidly decay with $r$~\cite{Sigrist96}. We confirmed this using ED calculations by introducing two impurities at varying distance $r$ along and transverse to the chains. The effective coupling between the two impurities states, shown in Fig.~\ref{DTNXfig4}, is found to be (i) controlled by the bare couplings $J_{a,b,c}$ and exponentially suppressed with $r/\lambda_{a,b,c}$ ($\lambda_{a,b,c} \simeq 2\xi_{a,b,c}$), and (ii) nonfrustrated, preserving the AF character of the underlying microscopic model. This nontrivial result paves the way to contemplate the level-crossing physics, unveiled by NMR, as the building blocks for a new type of ``order from disorder''~\cite{orderfromdisorder} mechanism. In close analogy with impurity-induced long-range order observed in various spin-gapped systems \cite{Bobroff09}, the nonfrustrated character of the effective coupling between localized two-level systems offers favorable conditions for the impurity states to eventually get \emph{ordered} at a low temperature, in stark contrast to previously reported Bose-glass physics in the same regime~\cite{Yu12}. Indeed, the ordering of impurity states could have been suspected from the previous theoretical study of a related model system~\cite{Nohadani05}, and the existence of a new (inhomogeneous) BEC phase centered on $H^*$ is now unambiguously confirmed by a new numerical study of DTNX~\cite{Dupont16}.

To conclude, our NMR study allowed us to precisely quantify the impurity states in DTNX, which are strongly localized and play a crucial role in the physics of the Bose-glass regime reported above $H_{\textrm{c2}}$. The system can be effectively described by two-level impurity states whose pairwise interaction is finite, albeit exponentially suppressed with distance. We have thus built a basis to theoretically access the many-body correlations between impurities. This enabled a critical reexamination of the phase diagram of DTNX above $H_{\textrm{c2}}$: the related theoretical work \cite{Dupont16} determines the (doping-dependent) extension of the field range where an inhomogeneous, ordered, BEC-type phase, related to the level crossing, is replacing a true Bose-glass regime. Our work thus sets a microscopic basis and proper delimitations of the impurity-induced BG regime for all further studies.

\begin{acknowledgments}
We acknowledge fruitful discussions with M. A. Continentino. This work has been supported by the French ANR project BOLODISS  (Grant No. ANR-14-CE32-0018) and by R\'{e}gion Midi-Pyr\'{e}n\'{e}es. A. P-F. acknowledges support from the Brazilian agencies CNPq and FAPESP (Grant No. \mbox{2015-16191-5}).
\end{acknowledgments}


\begin{thebibliography}{99}
\bibitem{gapped?} The fact that the spin dynamics is in a certain $T$ range dominated by gapped excitations does not exclude the possibility of existence of nongapped excitations which may show up only at much lower $T$.
\bibitem{note} There are two different Cl sites in each $J_c$ bond, and only one of these is replaced by doped Br. Therefore, $x$-percent doping modifies $2x$ bonds \cite{Yu12}.
\end{thebibliography}
\end{document}